\documentstyle[twocolumn,prb,aps,floats,epsf,lscape,grafik]{revtex}
\begin{document}
\draft
\twocolumn[\hsize\textwidth\columnwidth\hsize\csname@twocolumnfalse\endcsname

\title
{\bf  Suppression of superconductivity in high-$T_c$ cuprates due to 
nonmagnetic impurities: Implications for the order parameter symmetry}
\author{M. Bayindir and Z. Gedik}
\address{
Department of Physics,
Bilkent University,
Bilkent, 06533 Ankara, Turkey}
\date{\today} 
\maketitle

\begin{abstract}
We studied the effects of nonmagnetic impurities  on  high-temperature 
superconductors by solving the Bogoliubov-de Gennes equations  on a 
two-dimensional lattice via exact diagonalization technique  in a fully 
self-consistent way. We found that $s$-wave 
order parameter is almost unaffected by impurities at low  concentrations 
while $d_{x^2-y^2}$-wave  order parameter  exhibits a strong linear 
decrease with impurity concentration.  We evaluated the critical impurity 
concentration $n_i^c$ at which superconductivity ceases to be 0.1 which 
is in good agreement with experimental 
values. We also  investigated  how the orthorhombic nature of the crystal
 structure affects the suppression of superconductivity  and found that 
anisotropy induces an additional $s$-wave component. Our results support 
 $d_{x^2-y^2}$-wave  symmetry for tetragonal and $s+d_{x^2-y^2}$-wave 
symmetry  for orthorhombic structure.  
\end{abstract}
 
\vskip1pc]
 
\narrowtext
\newpage

Symmetry of the order parameter (OP)  in high-$T_c$ cuprates is 
important both for understanding the mechanism of superconductivity 
and  also for technological applications\cite{turk96}. For example, 
$d$-wave  symmetric OP effectively refutes phonon mechanism and for
a device made of a $d$-wave  superconductor having no gap in energy spectrum,
no refinement would get rid of the dissipation at low frequencies, 
even at low temperatures. 3d metal (Zn, Ni, Al, Ga, Fe, ...)  atom 
substitution for  Cu atoms  in high-$T_c$ cuprates may  identify symmetry 
of the OP \cite{acqu96}.  It is a well known fact that for conventional
superconductors having isotropic order parameter, nonmagnetic impurities 
with small concentrations have no effect  on  critical 
temperature\cite{lyto57,chan59,ande59,abri59} while magnetic impurities act 
as strong pair breakers, and as a result of this superconductivity is 
suppressed  very rapidly \cite{reif62,abri61,skal64}. On the other hand,
nonmagnetic impurities are very effective  in  anisotropic  
superconductors\cite{kada63,hohe64,abri93}.  For a pure  superconductor,
anisotropy leads to increase in $T_c$ \cite{kada63,pokr61,vall95}  and 
the critical temperature suppression rate with increasing impurity 
concentration  is proportional to  the strength of  
anisotropy\cite{kada63,hohe64,abri93}. Unlike the conventional 
superconductors, in hole-doped \cite{note1} high-$T_c$ cuprates  both 
magnetic (Ni) and nonmagnetic (Zn)  impurities suppress $T_c$ very
effectively.

Dependence of the superconducting properties (critical temperature, order 
parameter, density of states, ...)  on impurity or point defect concentration
is a subject of ongoing research. So far, most of the experiments  have 
been performed   to investigate effects of Zn and Ni  substitution in 
YBa$_2$Cu$_3$O$_{7-\delta}$  compounds\cite{xiao88,jaya88,chie91,rose92,brid93,semb94,kimj94,jano94,bonn94,ulm95,soer95,mend95,walk95,zago95,zhen96,fuku96,bern96,nach96,tail97,wang98,spec93,rao95,odag95,ulm96}. 
Alternatively, disorder can also be introduced  by  creating defects with
ion  irradiation \cite{giap94,jack95,eles96,tolp96,moff97}, but  
in this case affected region  is often uncontrollable.  

In spite of  the complexity of the high-$T_c$ cuprates (boundaries, defects, 
...), insufficient control of the actual impurity or point defect 
concentration,  solubility, and homogeneity of the distribution of the 
dopants which may lead to contradictory data, we can summarize  
some of the experimental results  as follows:

\begin{table}[t]
\caption{The critical temperature  $T_{c0}$ and the initial drop 
$\chi=[(T_{c0}-T_c(x))/T_{c0}]/x$  in various Zn doped YBCO compounds.
 In constructing the table, we used $T_c(x)$ values at $x<0.04$ for 
which $\chi$ is almost $x$ independent.}
\label{tablo1}
\begin{tabular}{llll} 
Material &  $ T_{c0}$ [K]&  $\chi$ &  Reference \\ 
\tableline
YBa$_2$(Cu$_{1-x}$Zn$_x$)$_3$O$_{7}$		& 92 & -13 &\ref{xiao88}\\ 
YBa$_2$(Cu$_{1-x}$Zn$_x$)$_3$O$_{7-\delta}$	& 90 &-12.3&\ref{chie91} \\ 
YBa$_2$(Cu$_{1-x}$Zn$_x$)$_3$O$_{7}$		&90 &-10.5&\ref{rose92}\\ 
YBa$_2$(Cu$_{1-x}$Zn$_x$)$_3$O$_{7}$		& 92 &-12.3&\ref{jano94}\\ 
YBa$_2$(Cu$_{1-x}$Zn$_x$)$_3$O$_{7}$		&92 &-15&\ref{ulm95}\\ 
YBa$_2$(Cu$_{1-x}$Zn$_x$)$_3$O$_{7}$		&87 &-8.7&\ref{zago95}\\ 
YBa$_2$(Cu$_{1-x}$Zn$_x$)$_3$O$_{6.9}$		&93.6 &-6.8&\ref{tail97}\\ 
YBa$_2$(Cu$_{1-x}$Zn$_x$)$_3$O$_{6.9}$		&93 &-15 & \ref{wang98} 
\end{tabular}
\end{table}

\begin{itemize}
\item  For  YBa$_2$(Cu$_{1-x}$Zn$_x$)$_3$O$_{7-\delta}$  compounds, at small 
concentrations, $x < 0.04$,  Zn  ions occupy preferably Cu-sites in the 
CuO$_2$ planes, however for  $x > 0.04$ the substituent starts to occupy 
Cu-sites in the chain \cite{xiao88,rose92,vill94,hodg94,klug96}. Since this 
compound  has two planes and one chain in a unit cell, for  $x < 0.04$ the 
actual (effective)  impurity concentration $n_i$, i. e. the number of 
impurities per unit cell per CuO$_2$ plane,  becomes $3x/2 $.
\item The critical temperature decreases linearly  with increasing impurity 
concentration in substitution \cite{xiao88,zago95} (for $x > 0.04$ the drop 
 rate decreases due to partial occupancy of Zn at chain sites) and point
 defect concentration irradiation\cite{moff97} experiments. 
\item Due to orthorhombicity  of  YBCO material,  CuO$_2$ planes exhibit an
  anisotropic  behavior  and admixture of  $d$- and $s$-wave is 
possible\cite{wang98,baso95,klei96,kouz97}.
\item Zn substitution does not alter the carrier concentration in 
 CuO$_2$ planes \cite{bern96}.
\end{itemize}

On the theoretical side, the existing pair-breaking models overestimate the 
suppression of  critical temperature and predict  an increasing slope with 
increasing impurity concentration\cite{kim95,radt93,arbe94,fehr96}  which 
contradicts the observed linear  dependence of $T_c$  on $n_i$. The effects
 of neither magnetic (Ni)  nor  nonmagnetic (Zn) impurities on the 
superconducting properties of the cuprates  have been  explained clearly. 
Main reason for the  discrepancy between theory and experiments   is 
that the conventional  Abrikosov-Gor'kov  (AG) -type  pair-breaking models
 ignore position dependence of the order parameter  near impurity sites. 
Recently, Franz  and his coworkers \cite{fran97}, and Zhitomirsky and 
Walker \cite{zhit98} argued that spatial variation of the order parameter 
must be taken into account for short coherence length superconductors. 

In the present paper, we investigate  effects of nonmagnetic impurities on 
high-$T_c$ cuprates for both tetragonal and orthorhombic phases by solving 
the Bogoliubov-de Gennes (BdG) equations \cite{bogo58,dege89} in a fully 
self-consistent way.  In particular, we address possibility  of extracting
  OP symmetry by examining  effects of nonmagnetic impurities.  Our results 
support  $d_{x^2-y^2}$-wave pairing symmetry for tetragonal and 
$s+d_{x^2-y^2}$-wave for orthorhombic  structure. Possibility  of 
admixture of $s$- and $d$-wave symmetries  have already been proposed 
 in various experimental\cite{klei96,kouz97} and theoretical 
works\cite{emer94,vare98,mono96,kim95}. 

The BdG  equations on two-dimensional lattice have the following 
form\cite{dege89,mart98}

\begin{equation}
\sum_j \left(  
 \begin{array}{cc} 
H_{ij}  &\Delta_{ij} \\  \Delta^\star_{ij} & -H^\star_{ij} 
\end{array} \right) \left(   \begin{array}{c} u_n(j) \\ v_n(j) 
\end{array} \right)= E_n  \left(   \begin{array}{c} u_n(i) \\ v_n(i) 
\end{array} \right)\;,
\label{bdg}
\end{equation}
where $u_n(i)$ and $ v_n(i) $ are quasiparticle amplitudes at site $i$ with 
eigenvalue $E_n$, and $\Delta_{ij} $ is the pairing potential. 
The normal-state part of the Hamiltonian can be written as 

\begin{eqnarray}
H_{ij}&=&(t_{ij}+U_{ij}n_{ij}/2)(1-\delta_{ij}) \nonumber \\
&&+(V_i^{\rm imp} -\mu+  U_{ii}n_{ii}/2)\delta_{ij}\;, 
\end{eqnarray}
where $t_{ij}$  is the hopping amplitude, $\mu$ is the chemical potential, 
$U_{ij}n_{ij}/2 $ and $U_{ii}n_{ii}/2 $ are  the Hartree-Fock potentials 
with on-site interaction $U_{ii}$ and off-site interaction $U_{ij}$,  
respectively. Finally $V_i^{\rm imp}$ is the impurity potential.  The 
pairing potentials are defined by 
\begin{equation}
\Delta_{ij} =-U_{ij} F_{ij}\;.
\end{equation}
The charge density $n_{ij}$ in the Hartree-Fock potentials and the anomalous
 density  $F_{ij}$ in the pairing potential are determined  from
\begin{equation}
n_{ij}=\sum_\sigma \langle \Psi_\sigma^ \dagger (i)\Psi_\sigma (j) \rangle \;,
\end{equation}
\begin{equation}
F_{ij}=\langle \Psi_\uparrow(i) \Psi_ \downarrow (j) \rangle \;,
\end{equation}
where $\sigma$ is spin index, and   $\Psi_\sigma^ \dagger (i)$ and 
$\Psi_\sigma (i)$ are related to the quasiparticle creation 
($\gamma^\dagger_{n\sigma}$) and annihilation ($\gamma_{n\sigma}$) operators  
\begin{equation}
\left[   \begin{array}{c}  \Psi_\uparrow (i)  \\ \Psi_\downarrow^ \dagger (i) 
\end{array} \right]=\sum_n \left[  \gamma_{n\uparrow} 
\left( \begin{array}{c} u_n(i) \\ v_n(i) 
\end{array} \right)+ \gamma_{n\downarrow} 
\left( \begin{array}{c} -v^\star_n(i) \\ u^\star_n(i) 
\end{array} \right) \right]\;,
\end{equation}
where $\gamma$ and $\gamma^\dagger$ satisfy the Fermi commutation relations. 
 The self-consistency  conditions can be written  in terms of $u_n$, 
 $v_n$, and $E_n$
\begin{eqnarray}
n_{ij}=2\sum_n u^\star_n(i) v_n(j) f(E_n) \nonumber \\
 + v_n(i) v^\star_n(j) [1-f(E_n)]\;, \label{self1}
\end{eqnarray}
\begin{eqnarray}
F_{ij}=\sum_n  u_n(i) v^\star_n(j) [1-f(E_n)] \nonumber \\
 -  v^\star_n(i) u_n(j) f(E_n)\;,\label{self2}
\end{eqnarray}
where $f(E_n)=1/[\exp{(E_n/k_BT)}-1]$ is the Fermi distribution function.

\bfig{t}\ff{0.45}{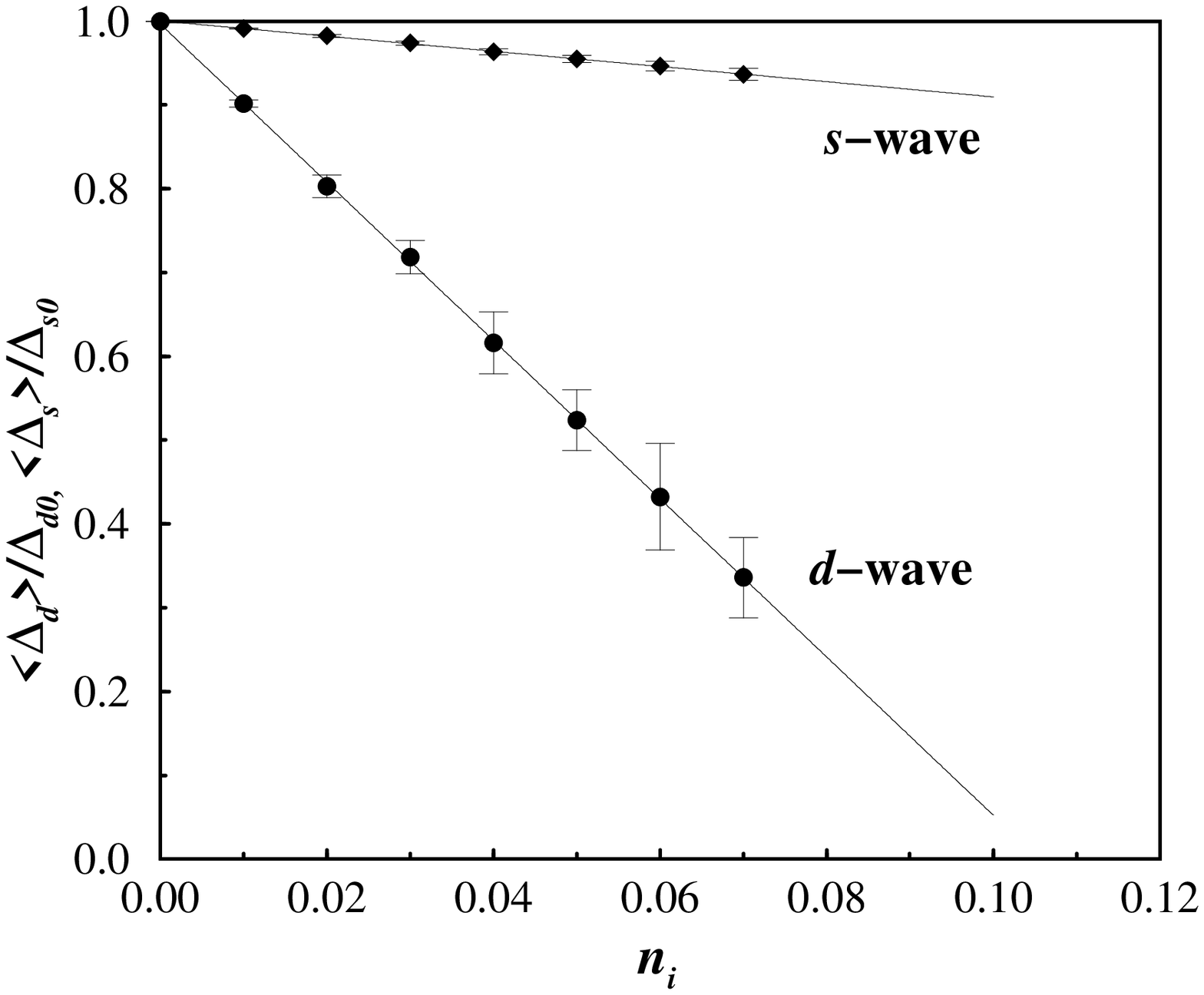}\efig{Normalized $s$- and $d$-wave order parameters,
 $\langle \Delta_{d} \rangle / \Delta_{d0}$ and 
 $\langle \Delta_{s} \rangle / \Delta_{s0}$, versus  impurity concentration
 $n_i$ for tetragonal structure. $\Delta_{d0}$  and  $\Delta_{s0}$  are 
 the magnitudes of the order parameters in the absence of the impurities,
 and $\langle \cdots \rangle$ is taken over 20 different  impurity 
ditributions.  Solid lines represent the best linear fit to the data.}{fig1}

We solve the BdG equations on a 20x20 square lattice (hence, we diagonalize 
a 1600x1600 matrix) with periodic boundary conditions by exact 
diagonalization technique using IMSL subroutines. After choosing a 
suitable initial guess for OP, we solve Eq.\ (\ref{bdg}). Next,  we calculate
 the new charge density $n_{ij}$ and anomalous density $F_{ij}$ via 
Eqs.\ (\ref{self1}) and (\ref{self2}) and iterate this procedure until 
a reasonable convergence is achieved.  The BdG equations are solved 
self-consistently. Self-consistency conditions [Eqs.\ (\ref{self1}) and
 (\ref{self2})] lead to 10 separate equations. The first   five [obtained from
 Eq.\ (\ref{self1})] renormalize on-site energies and hopping matrix 
elements while the last five  [obtained from Eq.\ (\ref{self2})] affect 
 the on-site and nearest-neighbor  interaction terms. Although,  the first 
five of these self-consistency conditions  can be neglected  for conventional 
superconductors  where $U/t \ll 1$,  for strong interaction case we should 
keep them, since they play important role  especially in the presence of 
impurities. The impurity potential $V_i^{\rm imp}$ is treated in the unitary 
limit, i.e. $V_i^{\rm imp} \gg t$, and  taken nonzero for randomly chosen 
lattice sites. 

We first solve  the BdG equations for tetragonal  case $t_x=t_y=t$ where 
$t_x$ and $t_y$  are nearest-neighbor hopping amplitudes along $x$ and $y$ 
directions, respectively. For $s$-wave OP  symmetry  we assume that on-site 
(attractive) interaction is $U_{ii}=-1.7t$ and there is no nearest-neighbor
 interaction.  In the case of $d$-wave  OP symmetry    on-site (repulsive)  
interaction is $U_{ii}=1.4t$ and  nearest-neighbor (attractive) interaction 
is $U_{ij}=-1.4t$. With this choice  of parameters we fix chemical potential
 $\mu$  so that  band filling factor $\langle n\rangle\simeq 0.8$ and  zero 
temperature coherence length $\xi_0 \simeq 4a$. These values are in good 
agreement with the commonly accepted  experimental values.

Figure~1 shows that, at low impurity concentrations $s$-wave OP symmetry is 
almost unaffected  by  the impurities or point defects. Although we use 
several on-site interaction values by keeping the band filling factor and 
zero temperature coherence length constant, we don't get any qualitative 
change. This result is consistent with Anderson theorem\cite{ande59} 
and AG theory\cite{abri61}. However, experimental data for high-$T_c$ 
cuprates  exhibit a much stronger suppression of superconductivity with 
increasing  disorder. 

\bfig{t}\ff{0.45}{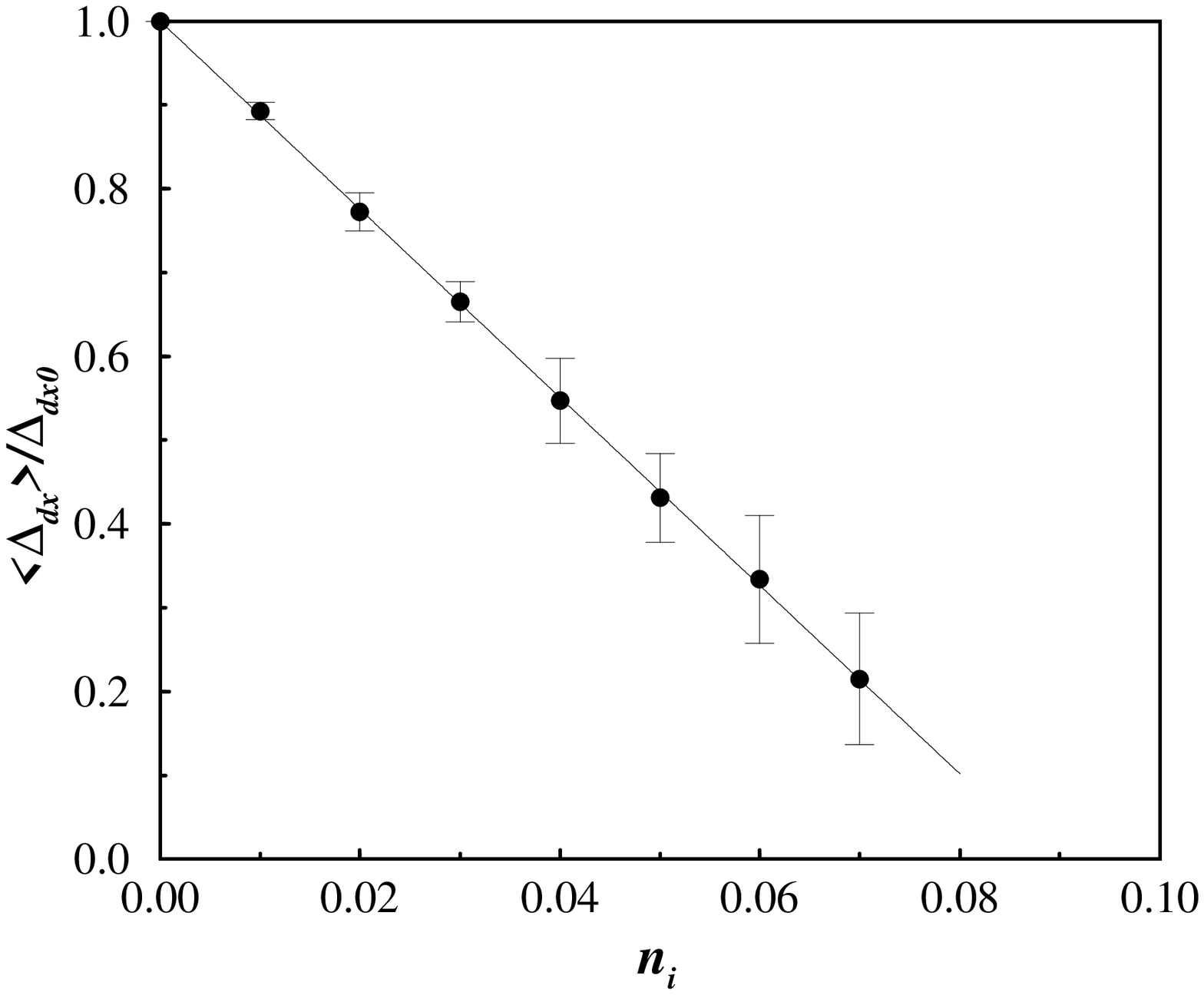}\efig{Normalized order parameter versus  impurity
 concentration  for orthorhombic structure. $\Delta_{dx0}$  is $x$ component
  of the order parameter in the absence of  impurities. Solid line represents
 the best linear fit to the data.}{fig2}

On the other hand, our $d$-wave calculations give results similar to the 
 behavior observed in experiments. For $d$-wave symmetry  we find that 
on-site pairing potential is negligibly small. In Fig.~1, $\Delta_d$ is 
amplitude of the nearest-neighbor pairing potential. We obtain a linear 
decrease in  the mean OP, which  is assumed to be proportional  to the 
critical temperature $T_c$ \cite{note2}, and  the slope of the straight 
line is in good agreement with the experimental data summarized in Table~I. 
The critical impurity or point defect concentration $n_i^c$ at which 
superconductivity ceases is also near to experimental value $\simeq 0.1$. 
In comparing  our results with experimental data we should  keep the 
following point in our mind. For $x < 0.04$, substitutional impurities  
go to preferentially CuO$_2$ planes\cite{xiao88,rose92,vill94,hodg94,klug96} 
 and hence the actual  concentration  is $3x/2$. However for higher  
concentrations some of the Zn atoms occupy the chain sites, and in this 
case we can not  relate the in  plane  concentration to the actual one. 
Therefore, we used  the initial points, i. e.   $x < 0.04$, to evaluate 
the initial drop in Table~I. To obtain the experimental value for  
critical impurity concentration $n_i^c$, we extrapolated the linear parts
 of the experimental curves to intersect the impurity concentration axes. 

Similar equations have already been solved by Xiang and Wheatley\cite{xian95}, 
however our additional self-consistency conditions and choice of
 parameters lead to a correct prediction for the critical 
impurity concentration.  It is important to note that we can not have a 
self-consistent solution of the OP for extended $s$-wave by using 
any physical values for the above model parameters. This fact was 
pointed out  by Wang and MacDonald\cite{wang95}, and they found that   
extended $s$-wave component is smaller than  $d$-wave component by about
two orders of magnitude.

\bfig{t}\ff{0.45}{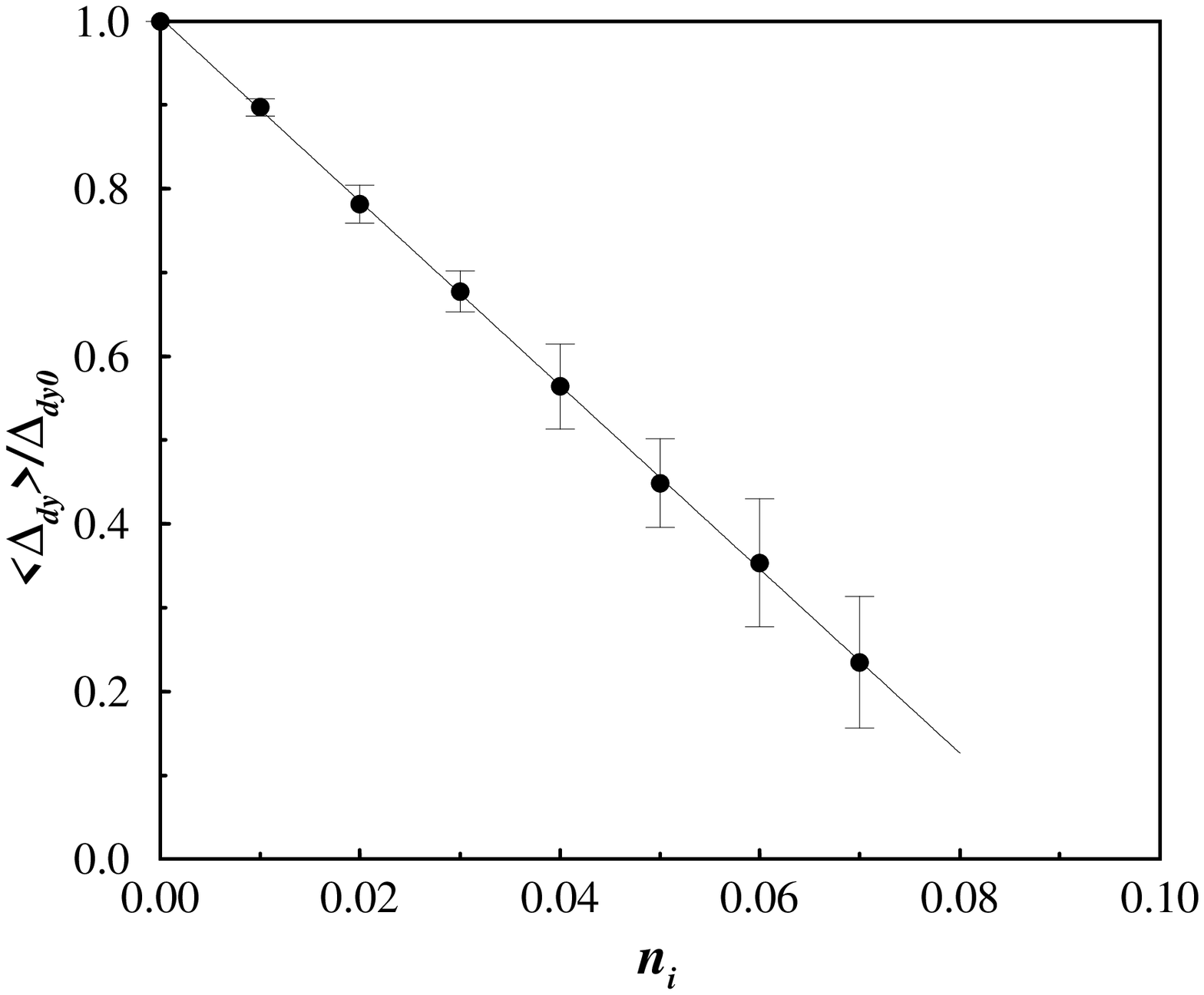}\efig{Normalized order parameter versus 
 impurity concentration for  orthorhombic structure. $\Delta_{dy0}$ 
 is   $y$ component of the order parameter in the absence of impurities.
 Solid line represents the best linear fit to the data.}{fig3}

When we introduce an orthorhombic distortion by taking $t_y=1.5t_x$, 
as suggested by experimental data\cite{wang98,baso95}, we observe 
that an $s$-wave component (of approximately ten percent of 
$d$-wave components) is induced. Figures  2 and 3 show the variation 
of $\Delta_{dx}$ and $\Delta_{dy}$ components, respectively. 
In the absence of disorder, $\Delta_{dy}/\Delta_{dx}\simeq 1.5$ 
with $\Delta_{dx}=0.082t$. With increasing disorder, the larger one, 
i.e. $\langle \Delta_{dy}\rangle$,  is suppressed faster. When we reach
  the critical impurity concentration both components vanish simultaneously. 
Moreover, $d$-wave components of the OP decrease linearly with $n_i$ 
 and vanish at $n_i\simeq 0.1$ as in the case of  pure $d$-wave symmetry. 
Here $\langle \cdots \rangle$ denotes averaging over 20 different impurity 
configurations.

\bfig{t}\ff{0.45}{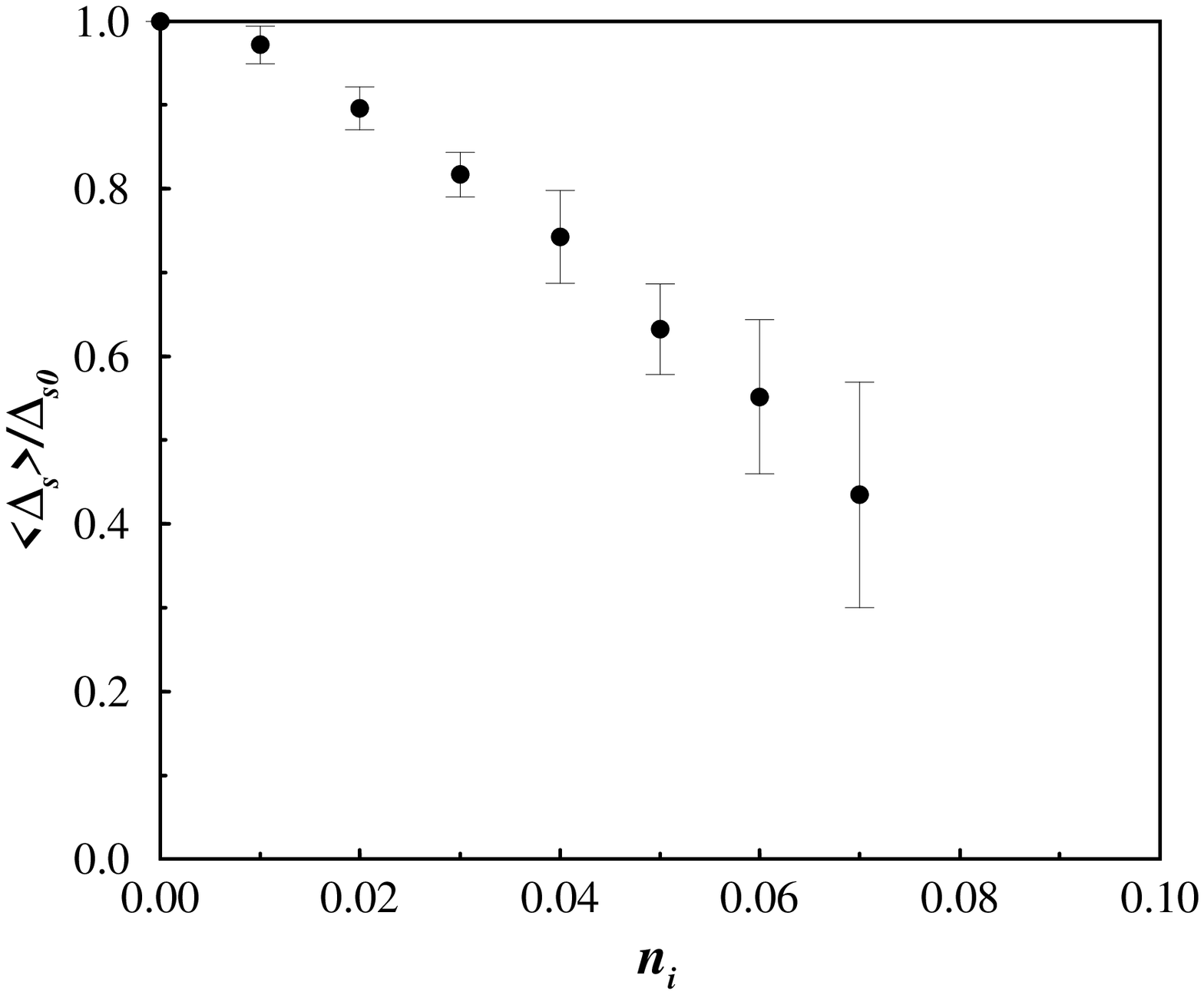}\efig{Normalized  $s$-wave  component of  the order
 parameter versus  impurity concentration  for orthorhombic  structure.
 $\Delta_{s0}$  is  the magnitude of the order parameter in the absence
 of  impurities. }{fig4}

As can be  seen from Fig.4, $s$-wave component also decreases with $n_i$, 
however while $d$-wave components exhibit a linear dependence on impurity 
concentration s-wave component shows a downward curvature similar to 
prediction of the AG theory.

In conclusion, we investigated the effects of nonmagnetic impurities and 
point defects within a BCS mean-field framework by means of BdG equations. 
 For tetragonal structure, we found out that  the observed suppression of 
superconductivity, when impurities are substituted or point defects are 
introduced,  can be explained only if the OP is  
$d_{x^2-y^2}$-wave symmetric.
 In  case of $s$-wave symmetry, superconductivity is almost unaffected 
by  disorder. When a slight anisotropy is introduced by distorting copper 
oxide planes from a square to rectangular lattice we observed that  a small
 amount of $s$-wave contribution is induced. For  both tetragonal and 
orthorhombic structures we evaluate the critical concentration at which 
superconductivity ceases to be very near to experimental value $\simeq 0.1$. 

{\it Acknowledgments}:

This work was supported by the Scientific and Technical Research 
Council of Turkey (TUBITAK) under grant No. TBAG 1736.


\begin{references}
\bibitem{note1} For electron-doped high-temperature superconductors, 
the situation is similar to BCS superconductors \cite{acqu96,maki98}.
\bibitem{note2} It is important to note that  near  $T_c$  the coherence 
length diverges and may become larger than the system size, therefore, 
 due to finite-size effects, determination of $T_c$ may be difficult.
\end{references}
\end{document}